\documentclass[conference=acm, camera-ready, savespace]{confpaper}

\usepackage[all=normal,wordspacing]{savetrees}

\bibliography{proceedings,paper}

\begin{document}

\title{Comparing the Effects of DNS, DoT, and DoH \\on Web Performance}
\renewcommand{\shorttitle}{Comparing the Effects of DNS, DoT, and DoH on Web Performance}

\author{Austin Hounsel}
\email{ahounsel@cs.princeton.edu}
\affiliation{%
    \institution{Princeton University}
}

\author{Kevin Borgolte}
\email{borgolte@cs.princeton.edu}
\affiliation{%
    \institution{Princeton University}
}

\author{Paul Schmitt}
\email{pschmitt@cs.princeton.edu}
\affiliation{%
    \institution{Princeton University}
}

\author{Jordan Holland}
\email{jordanah@cs.princeton.edu}
\affiliation{%
    \institution{Princeton University}
}

\author{Nick Feamster}
\email{feamster@uchicago.edu}
\affiliation{%
    \institution{University of Chicago}
}


\abstract{%
  Nearly every service on the Internet relies on the Domain Name System (DNS), which translates a human-readable name to an IP address before two endpoints can communicate.
  Today, DNS traffic is unencrypted, leaving users vulnerable to eavesdropping and tampering.
  Past work has demonstrated that DNS queries can reveal a user's browsing history and even what smart devices they are using at home.
  In response to these privacy concerns, two new protocols have been proposed: DNS-over-HTTPS (DoH) and DNS-over-TLS (DoT).
  Instead of sending DNS queries and responses in the clear, DoH and DoT establish encrypted connections between users and resolvers.
  By doing so, these protocols provide privacy and security guarantees that traditional DNS (Do53) lacks.

  In this paper, we measure the effect of Do53, DoT, and DoH on query response times and page load times from five global vantage points.
  We find that although DoH and DoT response times are generally higher than Do53, \emph{both protocols can perform better than Do53 in terms of page load times}.
  However, as throughput decreases and substantial packet loss and latency are introduced, web pages load fastest with Do53.
  Additionally, web pages successfully load more often with Do53 and DoT than DoH.
  Based on these results, we provide several recommendations to improve DNS performance, such as \emph{opportunistic partial responses and wire format caching}.
}

%


\maketitle

\section{Introduction}\label{sec:intro}
The Domain Name System (DNS) underpins nearly all Internet communication; DNS queries map human-readable domain names to corresponding IP addresses of Internet endpoints.
Because nearly every Internet communication is preceded by a DNS query, and because some applications may require tens to hundreds of DNS queries for a single transaction, such as a web browser loading a page, the performance of DNS is paramount.
Many historical DNS design decisions and implementations (e.g., caching, running DNS over UDP instead of TCP) have thus focused on minimizing the latency of each DNS query.

In the past several years, however, DNS privacy has become a significant concern and design consideration.
Past research has shown that DNS queries can reveal various aspects of user activity to eavesdroppers, including the web sites that a user is visiting~\cite{zhu2015connection}.
As a result, various efforts have been developed to send DNS queries over encrypted transport protocols.
Two prominent examples are DNS-over-TLS (DoT) and DNS-over-HTTPS (DoH).
In both cases, a client sends DNS queries to the resolver over an encrypted transport (TLS), which relies on the Transmission Control Protocol (TCP).

The use of encrypted transports makes it impossible for passive eavesdroppers to observe DNS queries on a shared network, such as a wireless network in a coffee shop.
These transports also allow clients to send encrypted DNS queries to a third-party recursive resolver (e.g., Google or Cloudflare), preventing a user's ISP from seeing the DNS queries of its subscribers.
As such, from a privacy perspective, DoT and DoH are attractive protocols, providing confidentiality guarantees that DNS previously lacked.

On the other hand, encrypted transports introduce new performance costs, including the overhead associated with TCP and TLS connection establishment, as well as additional application-layer overhead.
The extent of these performance costs is not well understood.
An early preliminary study by Mozilla found that queries with DoH are only marginally slower than conventional DNS over port 53 (Do53)~\cite{mozilla-experiment}.
However, Mozilla only measured query response times, which does not reflect the holistic end-user experience.

In this paper, we measure how encrypted transports for DNS affect end-user experience in web browsers.
We find that DNS queries are typically slower with encrypted transports.
Much to our surprise, however, we discovered that using DoT and DoH can result in \emph{faster} page load times compared to using Do53.
When exploring the underlying reasons for this behavior, we discovered that encrypted transports have previously ignored quirks that significantly affect application performance.
For example, when DNS queries are sent over a lossy network, DoT and DoH can recover faster than Do53 because TCP packets can be retransmitted after 2x the round-trip-time latency to a recursive resolver.

On networks with sub-optimal performance however, these protocols begin to suffer because of their connection and transport overhead.
The relative costs and benefits of a particular DNS transport protocol and its implementation for DNS query response times and web page load times ultimately depend on the underlying network conditions.
This variability suggests that in some cases, clients (i.e., operating systems or browsers) might consider using different transport protocols for DNS based on their varying cost, performance, and privacy trade-offs.
Our findings also suggest easy improvements to stub resolver and browser DNS implementations.

In this paper, we make the following contributions:
\begin{itemize}
    \item \emph{We provide a performance study of Do53, DoT, and DoH from five global vantage points.}
  We measure query response times and page load times using popular open recursive resolvers, as well as resolvers provided by local networks.

\item \emph{We show that encrypted DNS transports can lead to faster page load times than unencrypted DNS.}
  We find that DNS query response times for DoT and DoH are generally slower than Do53.
  Surprisingly, on lossy network conditions, page load times can be \emph{faster} when using DoT and DoH instead of Do53.
  We attribute this behavior to differences in retransmission times between UDP and TCP. 

\item \emph{We give applicable insights to optimize DNS performance.}
  During our measurements, we observed behavior in DNS implementations that could be capitalized on for performance.
  Based on these insights, we propose two optimizations: wire-format caching and opportunistic partial responses.
\end{itemize}

\section{Background}\label{sec:background}
At a high level, the process for resolving domain names into IP addresses works in several steps.
A client \textit{queries} a recursive resolver (``recursor''), for example, ``what is the IP address for {\tt example.com}?''
The client has traditionally been a stub resolver, which is a lightweight process that manages DNS interactions with the global DNS infrastructure.
If the recursor does not have an answer for the domain name cached, it will issue the query on the client's behalf to upstream servers in the DNS hierarchy, including the root, TLD, and ultimately authoritative servers for a given domain.
Once the answer is returned to the recursor, the recursor caches the response and sends it to the client.

Due to the historical origins of the DNS, there are several privacy problems that were not originally considered~\cite{rfc7626}.
For example, DNS queries sent over port 53 (or ``Do53'') are typically unencrypted.
This means that any eavesdropper listening to traffic between the client and a recursor can see what queries the client is making.
Such information can be used to reveal personal information, such as browsing patterns and client device types, which can then be used to link user identity with user traffic.
While recursors themselves could also observe every query a client makes, recent protocols have been introduced to (at least) improve privacy for DNS traffic in transit between clients and DNS servers.

Hu et al. proposed DNS-over-TLS (or ``DoT'') in 2016 to prevent eavesdroppers from observing DNS traffic between a client and a recursor~\cite{rfc7858}.
It works largely similar to Do53, but the DNS traffic is sent over an established TLS connection, which means that it relies on TCP by default rather than on UDP.
Once the connection is established, all queries are encrypted by the transport sent over port 853.
Although DoT is relatively new, it has seen a significant increase in popularity since its introduction as some operating systems, such as Android, have started to use DoT opportunistically~\cite{android-dot}.

In 2018, Hoffman et al. proposed DNS-over-HTTPS to prevent on-path manipulation of DNS responses~\cite{rfc8484}.
DoH is similar to DoT, but uses HTTP as the transport protocol instead of TCP.
Wire format DNS queries and responses are sent using HTTP, and client applications and servers are responsible for translating between the application-layer messages and traditional DNS infrastructure.
An argument for DoH versus DoT has surrounded anti-censorship concerns, as DoH uses port 443 compared with port 853.
Oppressive regimes sometimes censor the Internet by dropping DNS traffic, but DoH requires a malicious network operator to drop all HTTPS traffic (on port 443) to prevent name resolution.

In this paper, we do not investigate the privacy or anti-censorship properties offered by each protocol.
Rather, we are focus on the effects that Do53, DoT, and DoH have on web performance and analyzing their respective costs and benefits.
We believe such measurements are necessary for users to make informed decisions about protocol choice for this crucial function of the Internet.

\section{Method}\label{sec:method}
In this section, we define our performance metrics, explain how we measure them, and describe our experiment setup.

\subsection{Metrics}
To understand how Do53, DoT, and DoH affect browser performance, we measure page load times and DNS query response times.
Page load times are gathered through Mozilla Firefox, and DNS query response times are gathered using a custom tool.

\subsubsection{Page Load Time}\label{sec:page-load}
We use Mozilla Firefox 67.0.1 in headless mode controlled by Selenium to visit a list of websites and measure page load times.
We record page load times by inspecting HTTP Archive objects (HARs), which can be collected after a page has finished loading~\cite{har-spec}.
In particular, we extract the \texttt{onLoad} timing from each HAR, which measures the elapsed time between when a page load began to when the \texttt{load} event was fired.
Our measurement suite is packaged as a Docker image to enable reproducible measurements, and to clear the browser's HTTP cache between page loads.

The \texttt{load} event is fired when a web page and all of its resources have completely loaded.
It is specified in the HTML Living Standard and has been implemented by all major browser vendors~\cite{mdn-load}.
It has also been used to measure page load times in previous web performance research~\cite{butkiewicz2011understanding, dhawan2012fathom}.
A similar event is \texttt{DOMContentLoaded}, which is fired when the HTML for a web page has been loaded and parsed by the browser.
However, unlike the \texttt{load} event, it does not include the time for downloading each object on the page, which is necessary to understand how DNS protocols affect page load times~\cite{mdn-domcontentloaded}.

Another metric is above-the-fold time (AFT), which represents the time it takes to download and render content that is initially viewable within the browser's dimensions.
The motivation for measuring AFT is that users may perceive a page load to have finished before all the objects have been rendered.
However, to measure AFT, we would need to visually record the start time and end time of rendering within the browser's dimensions for each page load~\cite{above-the-fold}.
Given the large-scale nature of our measurements, this would be too cumbersome to measure.

\subsubsection{DNS Query Response Time}\label{sec:dns_method}
To obtain precise, accurate DNS query response times, we built a tool with the getdns and libcurl C libraries to issue Do53, DoT, and DoH queries. We measure response times for each unique domain in the HARs that we collect.
Importantly, we do not cache DNS responses with our tool.

Getdns is a library that provides a modern API for making Do53 and DoT queries in various programming languages~\cite{getdns}.
To simulate Firefox page loads, we enabled connection reuse for DoT with an idle timeout of 10 seconds in order to amortize the TCP handshake and TLS connection setup.
Although Firefox does not currently support DoT within the browser, we believe this is a realistic setting, as it is the default timeout used by DoT stub resolvers such as Stubby.
We also ensure that all Do53 queries are made over UDP.

Libcurl is a library that allows developers to use cURL features in their applications~\cite{libcurl}.
It supports POST requests over HTTPS, which can be used to make DoH queries after adding the MIME type ``application/dns-message''.
To issue DoH queries, we also enabled connection reuse, and we sent the queries over HTTP/2, which is the recommended minimum HTTP version for DoH~\cite{rfc8484}.
We independently verified that Firefox uses HTTP/2 through a packet capture with mitmproxy and Wireshark~\cite{mitmproxy, wireshark}).

Although HARs also provide DNS query response times, we discovered during the course of our experiments that the timings for individual components, including DNS query response times, are inaccurate.
For example, we discovered that the first query that a HAR contains can show DNS timings of 0~ms, even in cases where it is impossible because we begin every browsing session with an empty cache.
This is the case because, depending on how a website issues HTTP redirects, the first query in the HAR is not actually the first query that the browser performed.
Instead, the browser might have performed a variety of other HTTP requests and DNS queries before, which may still be in-progress or already cached.

Interestingly, this peculiarity not only results in timings of 0~ms, but other values as well.
The browser may issue multiple requests to the same domain at different times through its thread pool, with the first one being redirected (thus, itself not being in the HAR, and the redirection target having a timing of 0~ms), and other requests made in between resolving the name of the domain for the domain's first request.
In turn, the subsequent requests can be answered from the cache that the first request populated.
However, the first request does not appear in the HAR.
Depending on when the requests are made, which depends on factors such as rendering time, the timings can take any value and shift the timings to the left.
This would even be the case if we would use the maximum of all values, because the first request that triggers resolving the domain may not be present in the HAR.

\subsection{Experiment Setup}\label{sec:setup}
To ensure that our results representative of diverse network configurations, we perform measurements across multiple recursors and vantage points.
In addition to performing measurements from our instances in their default network conditions, we emulate cellular performance by applying traffic shaping.
This also enables us to understand how Do53, DoH, and DoT perform under poor network conditions, e.g. high latency and packet loss.
We describe our hardware and software configuration, choices of recursors, vantage points, network conditions, and websites below.

\subsubsection{Hardware and Software}
We deployed Amazon EC2 instances with the \texttt{m5.2xlarge} hardware configuration and the Debian Buster operating system.\footnote{We considered using PlanetLab for our measurements, but ultimately decided to use Amazon EC2 because we felt that we would get better performance guarantees.}
Each instance includes 32~GB of RAM, a 3.1 GHz Intel Xeon Platinum Processor (8 vCPU cores), and 10 Gbps of network bandwidth~\cite{ec2-instance-types}.
The machines are connected over Ethernet, and they run a measurement suite designed to collect page load times as well as DNS query response times.\footnote{Our tools are available at \url{https://github.com/noise-lab/dns-measurement-suite}.}
We deploy our Docker image and DNS tool across all machines.
We left all network settings in their default values for Firefox 67.0.1, except when we enabled DoH by setting \texttt{network.trr.mode = 3}.
This forces all DNS queries initiated by Firefox to be sent over DoH~\cite{firefox-doh-prefs}.
Importantly, Firefox 67.0.1 disables EDNS Client Subnet by default for their DoH implementation and enables DNS pre-fetching.


\subsubsection{DNS Recursors and Transport Protocols}
We measure how the selection of a recursor and DNS transport affect browser performance.
As such, we chose three popular public recursors: Google, Quad9, and Cloudflare.
Each resolver offers public name resolution for Do53, DoT, and DoH.
We also use the local recursor provided to our Amazon EC2 instances at each vantage point.
However, these recursors only supports Do53, and not DoT or DoH.
Thus, these recursors serve as baseline for browser performance over Do53.

Do53 and DoH are natively supported in Firefox, the browser we use to drive our page load time measurements.
However, as of October 2019, DoT must be configured by using a stub resolver on a user's machine outside of Firefox.
For our page load time measurements, we use Stubby for DoT resolution, a stub resolver based on the getdns library~\cite{stubby}.
Stubby listens on a loopback address and responds to for Do53 queries.
All DNS queries received by Stubby are then sent out to a configured recursor over DoT.
We modify \texttt{/etc/resolv.conf} on our measurement systems to point to the loopback address served by Stubby.
This forces all DNS queries initiated by Firefox to be sent over DoT.

We note that our goal is to perform natural experiments by using popular recursors that end-users choose.
As such, we are not able to control the caches of the recursors between measurements.
To avoid biasing results due to network quiet and busy times, as well as the potential effect of a query warming the recursor's cache for subsequent queries from the other protocols tested, we randomize several aspects of the measurement suite.
First, for each run through the list of websites, we shuffle the order of websites prior to browsing.
Next, for each individual website, we randomize the order of DNS protocol as well as the DNS provider.

\subsubsection{Provider Networks}\label{sec:networks}
Our goal is to understand relationships between page load times, DNS performance, and network performance.
DNS performance is greatly affected by a client's Internet service provider (ISP), as their network configuration determines the paths the DNS traffic will use to reach a resolver (should the client opt to use a resolver that is hosted outside of the ISP network).
To gain a general understanding of how DoH, DoT, and Do53 perform over different networks, we measure response times and page load times from five vantage points around the world.
We use Amazon EC2 to launch instances located in Ohio \& California (United States of America), Frankfurt (Germany), Sydney (Australia), and Seoul (South Korea).

\subsubsection{Emulated Network Conditions}
We are also interested in web performance over networks that exhibit packet loss or high latency.
We believe it is important to simulate cellular performance as an increasing number of users are browsing the web on their phones.
Furthermore, organizations like Cloudflare have released mobile applications to force the operating system to use encrypted DNS transports.
We perform our measurements using the default network conditions for our instances and three emulated mobile network conditions.
We dedicate an EC2 instance for each network condition at all vantage points, for a total of 20 instances.

To emulated mobile network conditions, we first apply traffic shaping to emulate 4G mobile network performance.
We shape outgoing traffic with an additional latency of 53.3 ms and jitter set to 1 ms.
We also dropped 0.5\% of packets to mimic the loss that cellular data networks can exhibit.
We then shape our uplink rate to 7.44 Mb/s and our downlink rate to 22.1 Mb/s.
These settings are based on an OpenSignal report of mobile network experience across providers~\cite{opensignal}.
Second, we apply traffic shaping to emulate a lossy 4G network.
We use the same latency and jitter settings as 4G, but we increase the loss rate to 1.5\% of packets.
For the remainder of the paper, we refer to this network condition as "lossy 4G."
Finally, we apply traffic shaping to emulate 3G network performance by adding 150 ms or latency and 8 ms of jitter, along with 2.1\% packet loss and uplink and downlink rates of 1 Mb/s each.
While users in well-connected areas are less likely to experience 3G performance, it remains prevalent globally, particularly in developing regions.

\subsubsection{Websites}
We collect HARs (and resulting DNS queries) for the top 1,000 websites on the Tranco top-list to understand browser performance for the average user~\cite{le2019tranco} visiting popular sites.
Furthermore, we measure the bottom 1,000 of the top 100,000 websites (ranked 99,000 to 100,000) to understand browser performance for websites that are less popular.
We chose to measure the tail of the top 100,000 instead of the tail of the top 1 million because we found through experimentation that many of the websites in the tail of the top 1 million were offline at the time of our measurements.
Furthermore, there is significant churn in the tail of top 1 million, which means that we would not be accurately measuring browser performance for the tail across the duration of our experiment.

\subsection{Limitations}
Our research has some limitations that may affect the generalization of our results.
Nonetheless, we argue that our work will further the research community's understanding of how DNS affects user experience, and how various DNS stakeholders can improve it.
First, we perform our measurements exclusively on the Debian operating system, which means that its networking stack and parameters for networking algorithms will affect our measurements.
However, networking stacks are often heavily optimized, so we expect our results to generalize across operating systems.
Second, we rely on Mozilla Firefox to measure page load times, which means that its DNS-related code will influence our results.
Considering that web browsers are among the most used software today and also highly optimized for performance, we also expect our results to generalize across browsers.
Finally, we conduct our experiments from Amazon EC2 instances, which are located in data centers.
On one hand, this means that we are not able to generalize our results across other networks, e.g. residential ISPs.
On the other hand, Amazon EC2 enables us to understand how Do53, DoT, and DoH perform with a certain network type from five global vantage points.

\section{Measurement Results}\label{sec:results}
Our measurements were performed continuously from September 17th, 2019 through October 12th, 2019 using the setup described in \Fref{sec:method}.
We did not introduce delay between each successive page load or only perform page loads at certain times of the day.
In this section, we describe our measurement results for query response times and page load times, and analyze the protocols to understand the performance.
These results provide some insight into how a user's choice of networks, recursors, and protocols affect browsing experience.
Due to space constraints, we are unable to provide plots for each of our five vantage points.
Instead, we highlight our vantage points in Frankfurt and Seoul.

From Frankfurt, the average latency to the anycast addresses for Cloudflare, Quad9, and Google was 1.03ms, 1.42ms, and 1ms, respectively.
From Seoul, the average latency to the anycast addresses for Cloudflare, Quad9, and Google was 26.65ms, 1.95ms, and 30.22ms, respectively.
These measurements were obtained by sending ICMP pings to each recursor after each attempted page load.
Unfortunately, the Amazon EC2 recursors in each vantage point dropped ICMP pings, so we were unable to to measure the latency from our instances to the recursors.
Nonetheless, given that our measurements were conducted from Amazon EC2 instances, the average latency to an Amazon EC2 recursor from each vantage point is likely lower than Cloudflare, Quad9, and Google.

\subsection{DNS Query Response Time}\label{sec:frankfurt_resolution}

Intuitively, DNS query response time is the most critical metric when characterizing DNS performance, as web pages typically include many objects (e.g., images, JavaScript, frames, etc.), which all must have their underlying server names resolved to IP addresses.
Indeed, previous work has shown that DNS queries can cause performance bottlenecks on website page loads~\cite{wang2013demystifying}.
Accordingly, we begin our study with the response times for our network environments.

We note that Mozilla conducted a measurement study of DoH query response times in 2018 with Firefox Nightly users.
In their measurement study, they found that most queries were 6~ms slower than Do53 queries, and that DoH actually has faster response times than Do53 for the slowest queries~\cite{mozilla-experiment}.
However, Mozilla's experiment was limited to Cloudflare's DoH recursor, and they report no data for other recursors, like Quad9 and Google.
Furthermore, they only measure DoH, leaving out DoT entirely.

To fill these gaps and independently validate Mozilla's results, we designed our own experiment to measure response times for Do53, DoT, and DoH across different networks and recursors.
For each HAR file that we collected with our automated browser, we extracted all unique domain names.
We then measure the response time for each domain name through our own tool, which uses getdns for Do53 and DoT queries, and libcurl for DoH queries.

\Fref{fig:frankfurt_dns} shows CDFs for DNS response times from Frankfurt for the top 1,000 websites and the top 99,000-100,000 websites combined.
As expected, we find that Do53 performs better than DoT and DoH on for most queries across all recursors.
The overhead introduced by encrypted transports for DoT and DoH generally leads to an increase in response time.
Interestingly, we find that DoH is slightly faster than Do53 for the slowest queries across all public recursors.
For example, with Cloudflare Do53, the mean response time is ${\approx}$34ms, and the standard deviation is ${\approx}$347ms.
However, with Cloudflare DoH, the mean response time is ${\approx}$40ms, and the standard deviation is ${\approx}$94ms.
We posit that this can be attributed to HTTP caching of the DNS wire-format, which we discuss more in~\ref{sec:wire_format}.

Comparing DoT with DoH, we see differences between providers.
Cloudflare DoT and DoH appear to perform equally for the majority of queries, though DoH begins to outperform DoT for queries that take longer than ${\approx}$50ms.
Google DoT generally outperforms DoH for queries that take less than ${\approx}$100ms, above which DoH performs better.
Quad9 shows the largest range in terms of performance, with DoT queries experiencing long latencies compared to all other recursors and protocols.
Quad9's DoH recursor tends to perform better in comparison, but still lags behind their Do53 service.

\subsection{Page Load Time}\label{sec:frankfurt_pageload_diff}

\begin{figure*}[t!]
    \centering
    \begin{subfigure}[t]{0.3\textwidth}
        \includegraphics[width=\linewidth]{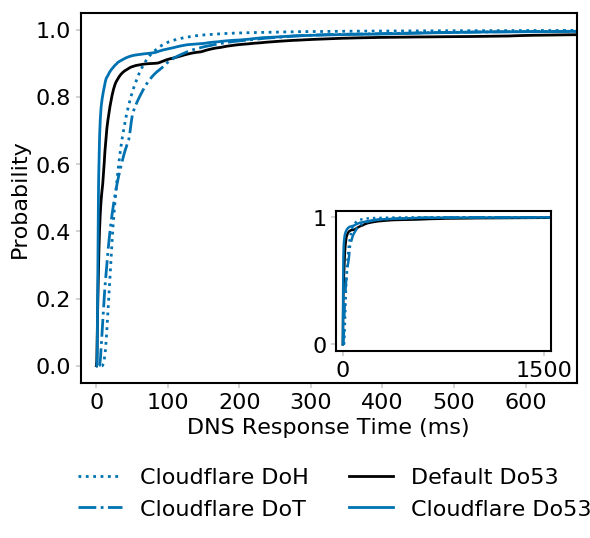}
        \caption{Cloudflare}
        \label{fig:frankfurt_dns_timings_cf}
    \end{subfigure}
    \hfill
    \begin{subfigure}[t]{0.3\textwidth}
        \includegraphics[width=\linewidth]{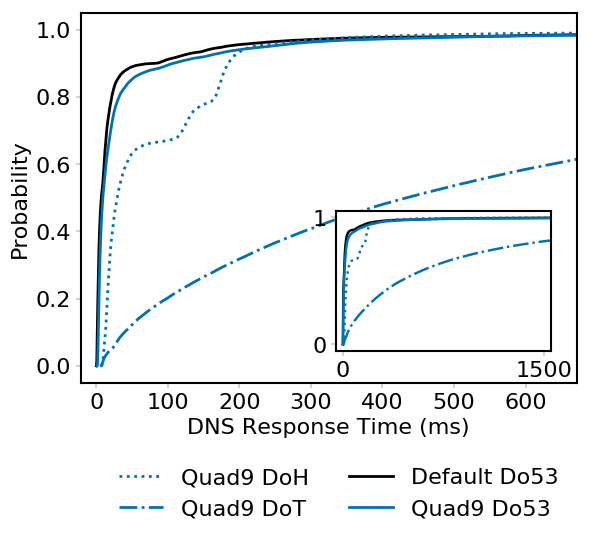}
        \caption{Quad9}
        \label{fig:frankfurt_dns_timings_quad9}
    \end{subfigure}
    \hfill
    \begin{subfigure}[t]{0.3\textwidth}
        \includegraphics[width=\linewidth]{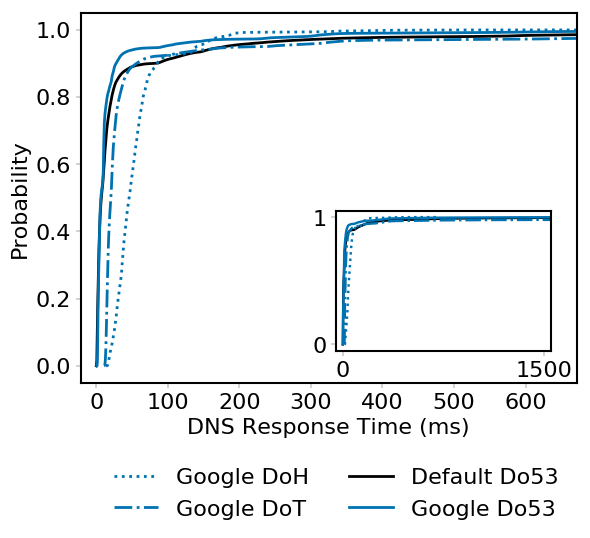}
        \caption{Google}
        \label{fig:frankfurt_dns_timings_google}
    \end{subfigure}
    \caption{Query response times for each provider from Frankfurt.}
    \label{fig:frankfurt_dns}
\end{figure*}
\begin{figure*}[t]
    \includegraphics[width=\textwidth]{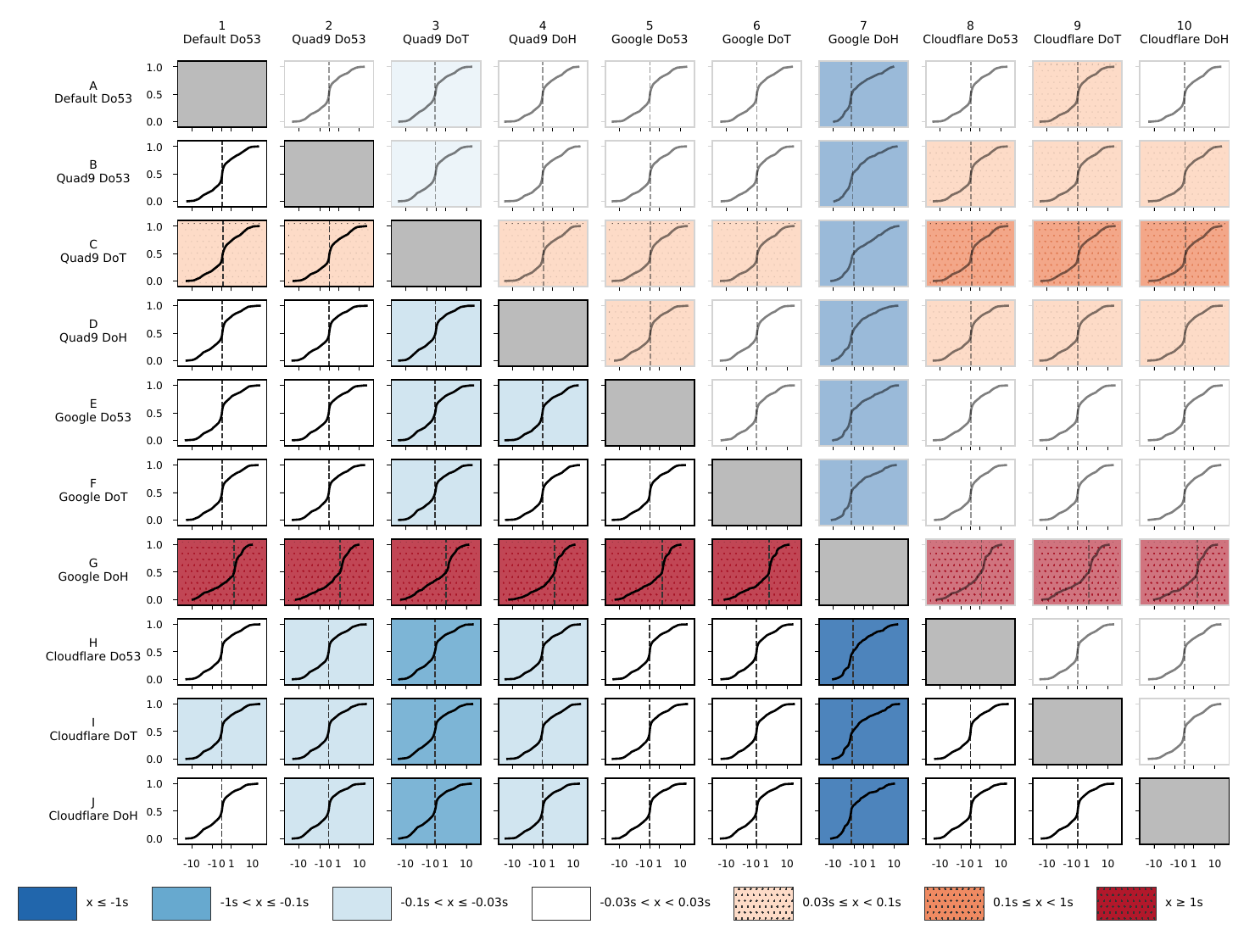}
    \caption{CDFs for differences in page load times between each configuration from Frankfurt.}
    \label{fig:frankfurt_pageload_diff}
\end{figure*}

Based on our results for query response times, we expect page load times to follow a similar pattern, with Do53 outperforming both DoT and DoH.
\Fref{fig:frankfurt_pageload_diff} shows CDFs for differences in page load times between each configuration when running our measurements from Frankfurt.
The vertical line on each subplot indicates the median for the CDF.
A median that is less than 0s on the x-axis means that the configuration (recursor, protocol) specified by the row title is faster than the configuration specified by the column title (indicated in blue hues).
Correspondingly, a median that is greater than 0s on the x-axis means that the configuration specified by the row title is slower than the configuration specified by the column title (indicated in red hues).
Finally, a median that is close to 0s (between -30ms and 30ms) indicates that row configuration and column configuration perform similarly.

Interestingly, for Cloudflare, each protocol finished within 30ms of each other for the median page load time.
These results stand in contrast to our expectation that page load times for DoT and DoH would be slower than Do53 due to additional latency for individual queries.
We posit that Cloudflare Do53, DoT, and DoH perform similarly in page load times because Firefox can resolve multiple names at once.
For Do53 and DoT, Firefox resolves names synchronously with a thread pool~\cite{firefox-thread-pool}.
Queries are sent via the operating system through through \texttt{getaddrinfo()})~\cite{dns-firefox-native-lookup}.
Furthermore, Firefox's DoH implementation is asynchronous, and it uses the browser's optimized HTTP/2 implementation~\cite{doh-firefox-trr-lookup, doh-firefox-trr}.
This means that DoH may be able to make up for its larger overhead compared to Do53 and DoT because page loads won't be blocked by synchronous queries if the thread pool is exhausted.

We find that Cloudflare Do53 and Google Do53 perform faster than the local Do53 recursor in median page load times.
We attribute this behavior to the caches of Cloudflare and Google more often containing the domain names we measured than the local recursor.
For example, as a CDN, Cloudflare is able to more quickly respond to DNS queries for domain names that they host than the other recursors~\cite{cloudflare-dns-overview}.
Cloudflare and Google also offer two of the most popular DNS services in the world, with 0.74\% and 9\% of users configuring their Do53 recursors, respectively.
This enables Cloudflare and Google to quickly respond to Do53 queries for a very large set of websites.
On the other hand, the local Do53 recursor was provided by Amazon for EC2 instances, which may not be used as often to query the domains of websites.

We also find that Google DoH performs significantly worse than all other DNS recursors or protocols from Frankfurt.
For example, when using Google DoH instead of Cloudflare DoH--\emph{the same website loads 1.35s slower in the median case}.
It may be the case that Google DoH's caching backend differs from their Do53 and DoT backends, which leads to longer page load times.
We note that as of October 2019, Google was in the process of migrating their DoH deployment to their production anycast address (8.8.8.8), and to fully support RFC 8484~\cite{google-doh-migration}.
During our experiments, we used the 8.8.8.8 anycast address and Google's production URI (\url{https://dns.google/dns-query}) to issue DoH queries, as advised in their documentation.

Similarly, Quad9 DoT performs worse in page load times than all recursors besides Google DoH, and a website loads 121ms faster using Cloudflare DoT over Quad9 DoT.
We offer several possible explanations.
For example, Quad9 DoT may not correctly cache responses, which leads to stacked normal distributions for the connection to the recursor.
This coincides with our data shown by Figure~\ref{fig:frankfurt_dns_timings_quad9}, in which only ${\approx}$20\% of Quad9 DoT queries completed in under 100ms.
Another possible explanation is that the recursor is trying to connect to authoritative nameservers via DoT, which fails and then triggers a retry via Do53.
Initially, when we disclosed our findings to Quad9, we did not receive an explanation. 
However, we were later informed that their DoT implementation was being changed.

\begin{figure*}[t!]
    \centering
    \begin{subfigure}[t]{0.3\textwidth}
        \includegraphics[width=\linewidth]{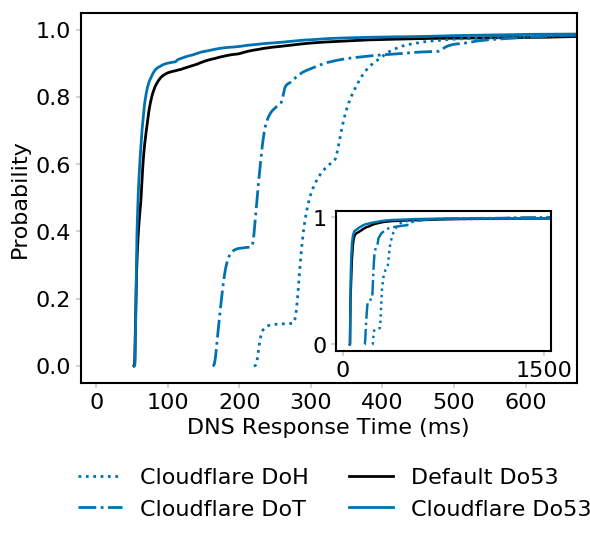}
        \caption{4G network}
        \label{fig:frankfurt_4g_dns}
    \end{subfigure}
    \hfill
    \begin{subfigure}[t]{0.3\textwidth}
        \includegraphics[width=\linewidth]{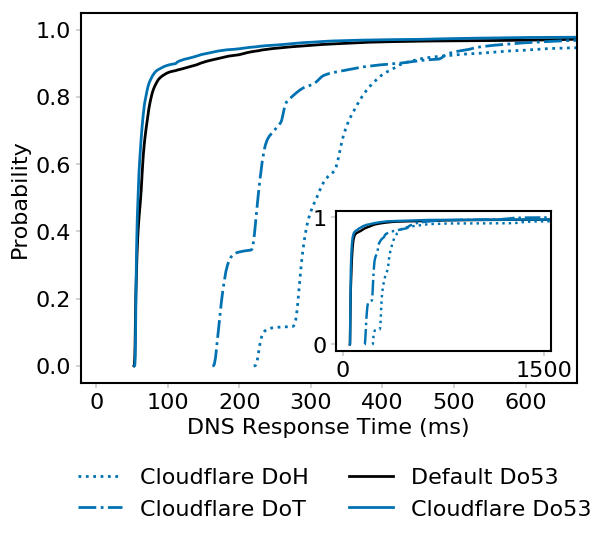}
        \caption{Lossy 4G network}
        \label{fig:frankfurt_4g_lossy_dns}
    \end{subfigure}
    \hfill
    \begin{subfigure}[t]{0.3\textwidth}
        \includegraphics[width=\linewidth]{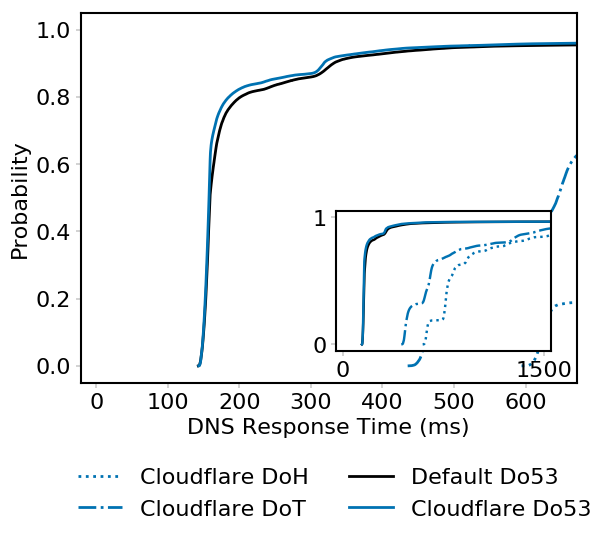}
        \caption{3G network}
        \label{fig:frankfurt_3g_dns}
    \end{subfigure}
    \caption{Query response times for Cloudflare across each protocol on three emulated networks}
    \label{fig:frankfurt_cellular_dns}
\end{figure*}

\subsection{Effect of Network Conditions}
We also study how network conditions affect query response times and page load times for Do53, DoT, and DoH.
Our results in \Fref{sec:frankfurt_resolution} and \Fref{sec:frankfurt_pageload_diff} are based on measurements conducted from a well-connected network in Frankfurt.
However, cellular network users in developing regions often access the Internet through networks with high latency and significant loss.
We expect such less-than-ideal conditions of these networks may significantly affect how Do53, DoT, and DoH perform.

\begin{table}[h!]
  \centering
  \footnotesize
  \rowcolors{1}{}{lightgray}
  \renewcommand{\arraystretch}{1.4}
  \begin{tabularx}{\columnwidth}{Xlrrr}
    & & \multicolumn{3}{c}{\textbf{Cloudflare}}\\
      \rowcolor{black}\cline{3-5}
    \rowcolor{white}\textbf{Connectivity} & \textbf{Status} & Do53 & DoT & DoH\\
    \arrayrulecolor{black}\hline
      \cellcolor{white}
        & Successful
        & 78.70\%
        & 78.65\%
        & 78.85\%
        \\
      \cellcolor{white}
        & Page-load Timeout
        & 7.48\%
        & 7.47\%
        & 7.21\%
        \\
      \cellcolor{white}
        & DNS Error
        & 9.51\%
        & 9.46\%
        & 9.90\%
        \\
      \cellcolor{white}
        & Selenium Error
        & 1.69\%
        & 1.74\%
        & 1.78\%
        \\
      \cellcolor{white}\multirow{-5}{*}{Default}
        & Other Error
        & 2.62\%
        & 2.67\%
        & 2.27\%
        \\
    \arrayrulecolor{gray}\hline
    \cellcolor{lightgray}
        & Successful
        & 80.02\%
        & 79.71\%
        & 78.61\%
        \\
      \cellcolor{lightgray}
        & Page-load Timeout
        & 7.86\%
        & 7.75\%
        & 7.22\%
        \\
      \cellcolor{lightgray}
        & DNS Error
        & 9.02\%
        & 9.00\%
        & 9.77\%
        \\
      \cellcolor{lightgray}
        & Selenium Error
        & 1.84\%
        & 1.67\%
        & 1.86\%
        \\
      \cellcolor{lightgray}\multirow{-5}{*}{4G network}
        & Other Error
        & 1.26\%
        & 1.87\%
        & 2.53\%
        \\
    \arrayrulecolor{gray}\hline
    \cellcolor{white}
        & Successful
        & 78.29\%
        & 78.13\%
        & 76.95\%
        \\
      \cellcolor{white}
        & Page-load Timeout
        & 8.24\%
        & 8.16\%
        & 8.01\%
        \\
      \cellcolor{white}
        & DNS Error
        & 9.95\%
        & 9.95\%
        & 10.76\%
        \\
      \cellcolor{white}
        & Selenium Error
        & 1.99\%
        & 1.96\%
        & 2.01\%
        \\
      \cellcolor{white}\multirow{-5}{*}{Lossy 4G network}
        & Other Error
        & 1.54\%
        & 1.80\%
        & 2.28\%
        \\
    \arrayrulecolor{gray}\hline
    \cellcolor{lightgray}
        & Successful
        & 28.10\%
        & 27.87\%
        & 20.06\%
        \\
      \cellcolor{lightgray}
        & Page-load Timeout
        & 60.02\%
        & 60.31\%
        & 41.32\%
        \\
      \cellcolor{lightgray}
        & DNS Error
        & 9.83\%
        & 9.76\%
        & 37.15\%
        \\
      \cellcolor{lightgray}
        & Selenium Error
        & 1.65\%
        & 1.54\%
        & 1.07\%
        \\
      \cellcolor{lightgray}\multirow{-5}{*}{3G network}
        & Other Error
        & 0.40\%
        & 0.51\%
        & 0.40\%
        \\
    \arrayrulecolor{black}\hline
  \end{tabularx}
  \caption{Successful website page-loads and error percentages for different network conditions when using Cloudflare's recursor from Frankfurt.}
  \label{tab:failure_cf_only}
\end{table}

\Fref{fig:frankfurt_4g_dns} and \Fref{fig:frankfurt_4g_lossy_dns} show CDFs for query response times with Cloudflare's recursor on an emulated cellular 4G network and an emulated lossy cellular 4G network.
We focus on Cloudflare's recursor because it performs better than Quad9 and Google (\Fref{fig:frankfurt_dns} and \Fref{fig:frankfurt_pageload_diff}).
On each emulated cellular network, Do53 outperforms DoT and DoH in terms of response time.
Interestingly, it appears that DNS timings on a cellular 4G and lossy cellular 4G network are similar, independent of the additional 1\% loss.

\Fref{fig:frankfurt_3g_dns} shows CDFs for response times for 3G network characteristics, which have higher loss, higher latency, and less bandwidth than 4G networks, and, in turn, we expect it affects DNS performance dramatically.
We find that DoT and DoH response times are substantially longer than Do53 response times.
The fastest DoT and DoH queries take ${\approx}$450ms and ${\approx}$600ms, respectively, where as the fastest Do53 queries take ${\approx}$150ms.
In fact, even the slowest DoH and DoT queries never close the latency gap to the slowest Do53 queries.

\begin{figure*}[t!]
  \centering
  \begin{subfigure}[t]{0.23\textwidth}
      \includegraphics[width=\linewidth]{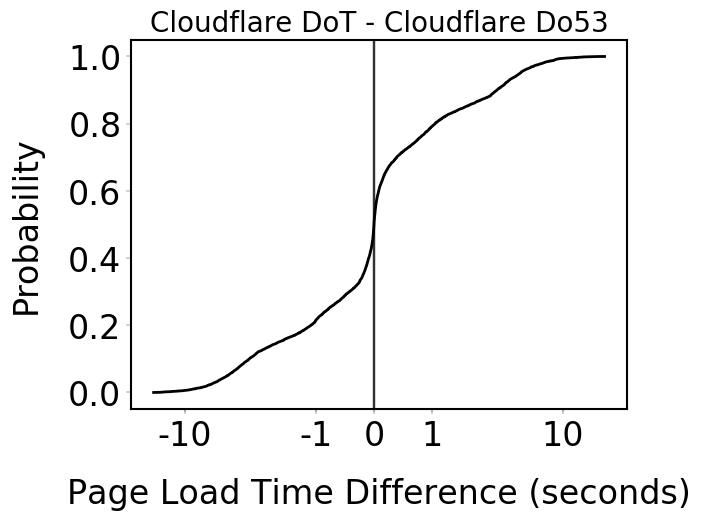}
      \caption{DoT - Do53, Frankfurt's default network}
      \label{fig:frankfurt_princeton_dot_dns}
  \end{subfigure}
  \hfill
  \begin{subfigure}[t]{0.23\textwidth}
      \includegraphics[width=\linewidth]{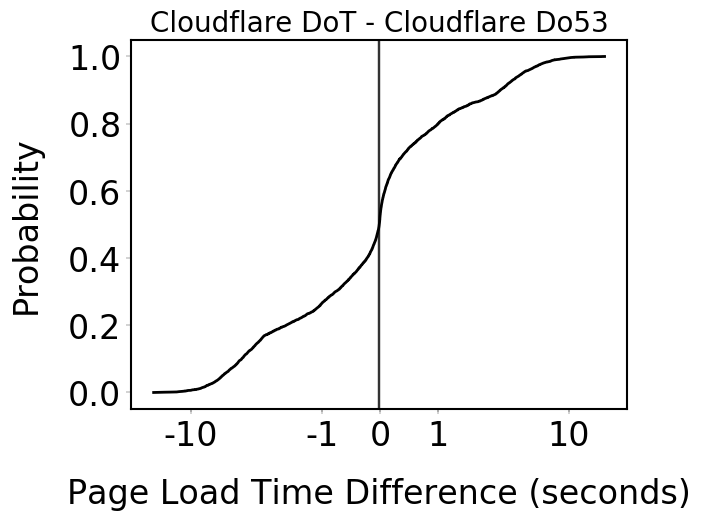}
      \caption{DoT - Do53, 4G network}
      \label{fig:frankfurt_4g_dot_dns}
  \end{subfigure}
  \hfill
  \begin{subfigure}[t]{0.23\textwidth}
      \includegraphics[width=\linewidth]{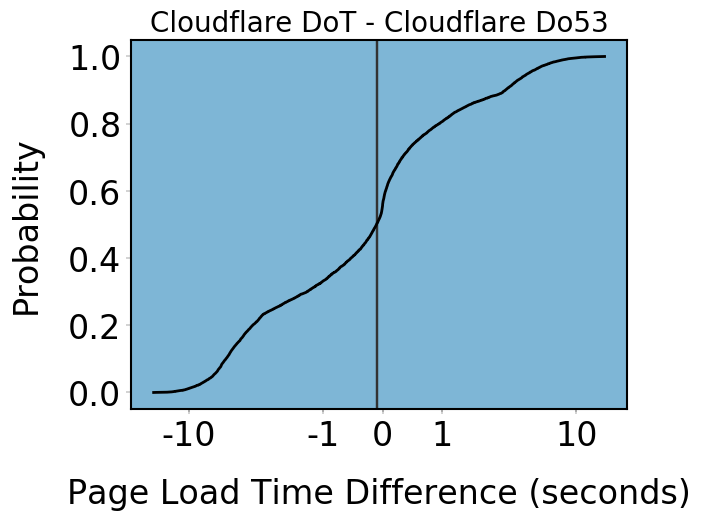}
      \caption{DoT - Do53, lossy 4G network}
      \label{fig:frankfurt_4g_lossy_dot_dns}
  \end{subfigure}
  \hfill
  \begin{subfigure}[t]{0.23\textwidth}
      \includegraphics[width=\linewidth]{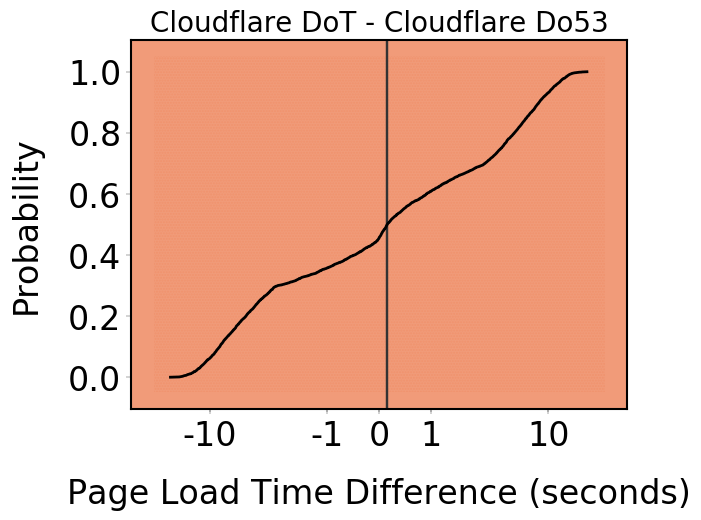}
      \caption{DoT - Do53, 3G network}
      \label{fig:frankfurt_3g_dot_dns}
  \end{subfigure}
\\
    \vspace{0.5cm}
  \begin{subfigure}[t]{0.23\textwidth}
      \includegraphics[width=\linewidth]{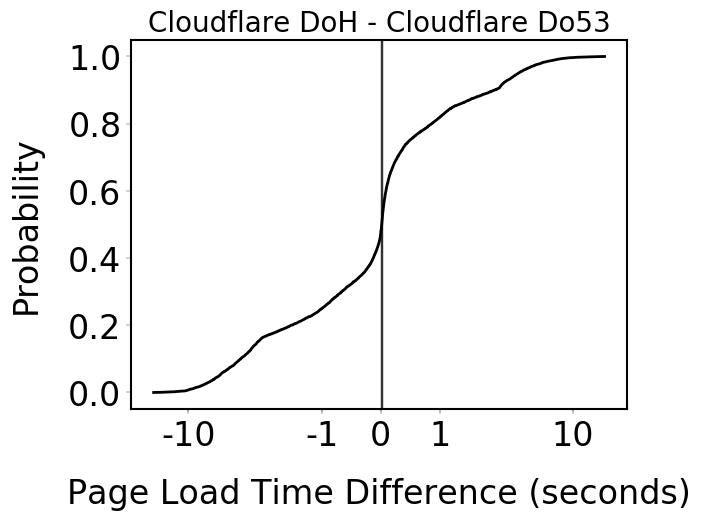}
      \caption{DoH - Do53, Frankfurt's default network}
      \label{fig:frankfurt_princeton_doh_dns}
  \end{subfigure}
  \hfill
  \begin{subfigure}[t]{0.23\textwidth}
      \includegraphics[width=\linewidth]{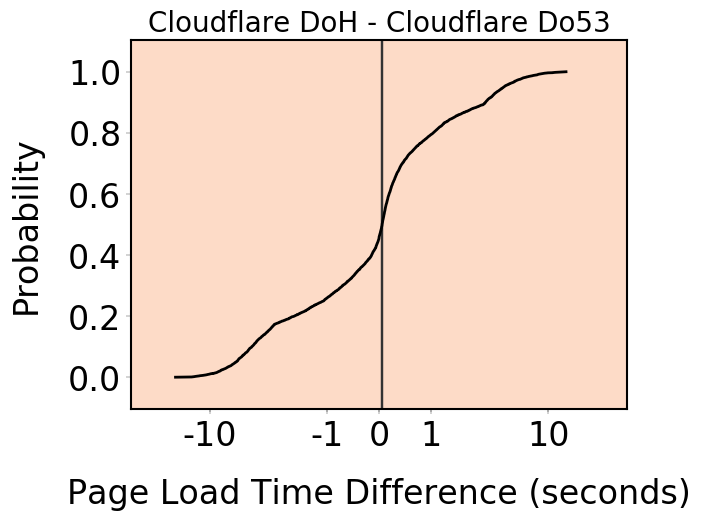}
      \caption{DoH - Do53, 4G network}
      \label{fig:frankfurt_4g_doh_dns}
  \end{subfigure}
  \hfill
  \begin{subfigure}[t]{0.23\textwidth}
      \includegraphics[width=\linewidth]{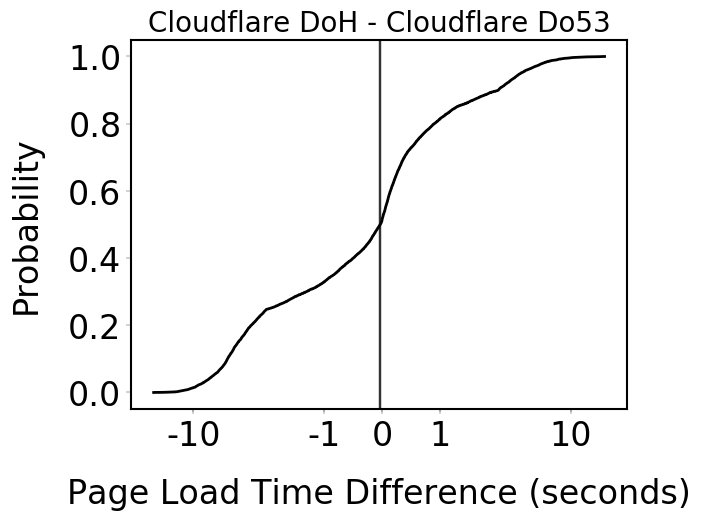}
      \caption{DoH - Do53, lossy 4G network}
      \label{fig:frankfurt_4g_lossy_doh_dns}
  \end{subfigure}
  \hfill
  \begin{subfigure}[t]{0.23\textwidth}
      \includegraphics[width=\linewidth]{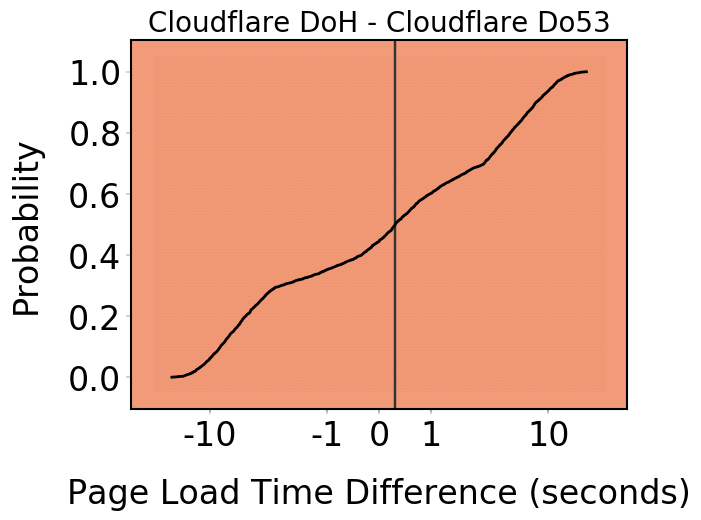}
      \caption{DoH - Do53, 3G network}
      \label{fig:frankfurt_3g_doh_dns}
  \end{subfigure}
\\
  \begin{subfigure}[t]{\textwidth}
      \includegraphics[width=\linewidth]{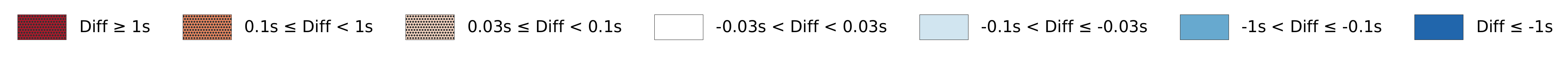}
      \label{fig:frankfurt_cf_pageloads_legend}
  \end{subfigure}
    \caption{Comparison of page load times between protocols and network conditions using Cloudflare's recursors from Frankfurt}
    \label{fig:frankfurt_cf_pageloads}
\end{figure*}

Based on the differences we observed in response times, we expected page load times on the emulated networks to be better with Do53 than with DoT or DoH.
\Fref{fig:frankfurt_cf_pageloads} compares page load times across all of our networks and protocols for Cloudflare's recursors.
Interestingly, on the 4G network, the median page load with DoT performs 11ms faster than Do53, and DoH performs 58ms slower.
On the lossy 4G network, DoT \textit{and} DoH are faster than Do53.
DoT performs 101ms faster than Do53, and DoH performs 33ms faster.

It may seem counter-intuitive that page loads using DoT and DoH perform these ways on the 4G and lossy 4G networks due to substantially longer queries (\Fref{fig:frankfurt_cellular_dns}).
However, the differences in how DNS timeouts are handled between TCP and UDP offer a possible explanation.
For example, the default timeout for Do53 queries in Linux is set to 5 seconds by \texttt{resolvconf}~\cite{resolvconf}.
For DoT and DoH, DNS packets may be retransmitted after 2x the round-trip-time latency to a recursor because of TCP.
If the round-trip time to a recursor is on the order of hundreds of milliseconds, then DoT and DoH will more quickly re-transmit dropped packets than Do53.

However, as throughput decreases and loss increases on a 3G network, DoT and DoH are no longer able to perform as well as Do53 concerning website page loads.
We believe this can be attributed to their higher overhead in bytes sent compared to Do53, which contributes to link saturation for most websites.
DoH also has a higher overhead than DoT, which leads to significantly slower page loads (\Fref{fig:frankfurt_3g_dot_dns} and \Fref{fig:frankfurt_3g_doh_dns}).
Furthermore, not only are more bytes are sent with DoT and DoH, but high latency and random packet loss significantly affect TCP performance~\cite{lakshman1997performance}. 

\Fref{tab:failure_cf_only} shows the prevalence and types of errors we encountered during our page load measurements.
Overall, we see that in lossier conditions, DoH experiences higher failure rates compared with Do53.
For instance, using the 3G settings, Cloudflare Do53 has ${\approx}$8\% less page load timeouts compared to Cloudflare DoH.
We also see that DNS errors spike for DoH in poor network conditions.
Conversely, DoT tends to maintain higher rates of success compared with DoH.
We note that there is a higher success rate in page loads with the 4G network condition compared to the default network condition.
It is not clear to us what caused this outcome.
We emphasize that our 4G, lossy 4G, and 3G network conditions were emulated; we did not perform measurements on real mobile networks.

\subsection{Trends Across Vantage Points}
Due to space constraints, we are unable to fully explore our results from other vantage points.
However, we observed that Cloudflare DoH and DoT were able to perform comparably to and sometimes better than Do53 on emulated cellular networks, regardless of the vantage point that was chosen.
In this section, we explore page load times on emulated network conditions in Seoul.

\begin{figure*}[t!]
  \centering
  \begin{subfigure}[t]{0.23\textwidth}
      \includegraphics[width=\linewidth]{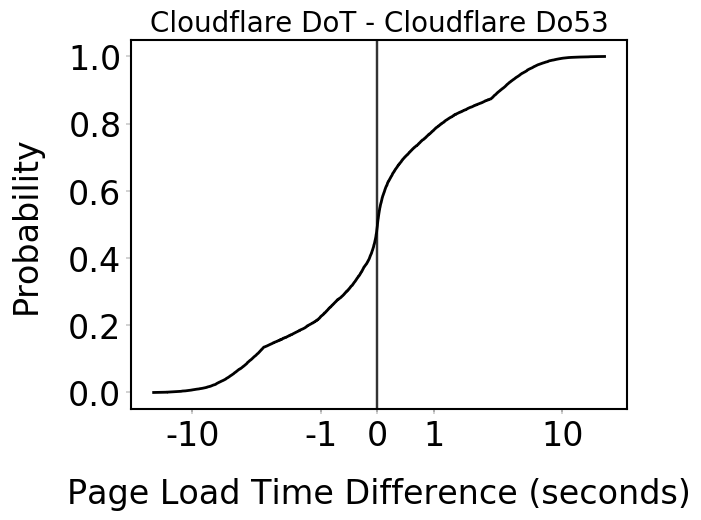}
      \caption{DoT - Do53, Seoul's default network}
      \label{fig:seoul_default_dot_dns}
  \end{subfigure}
  \hfill
  \begin{subfigure}[t]{0.23\textwidth}
      \includegraphics[width=\linewidth]{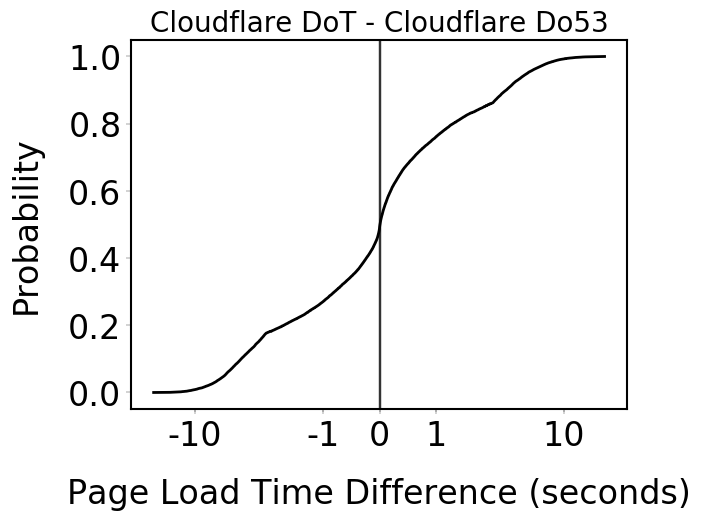}
      \caption{DoT - Do53, 4G network}
      \label{fig:seoul_4g_dot_dns}
  \end{subfigure}
  \hfill
  \begin{subfigure}[t]{0.23\textwidth}
      \includegraphics[width=\linewidth]{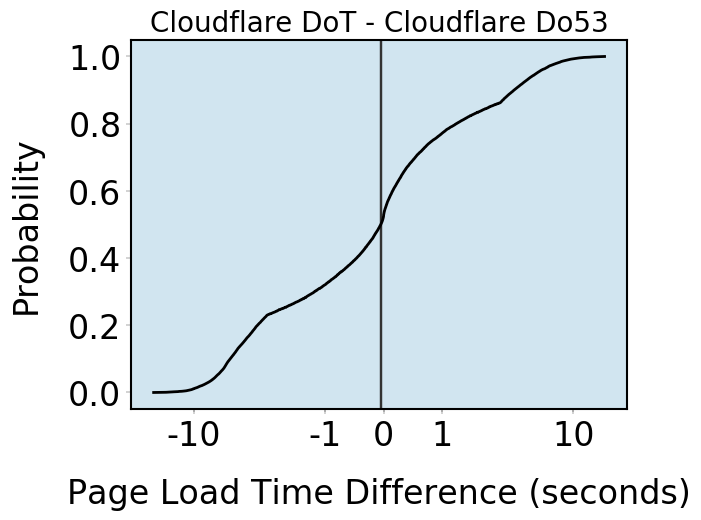}
      \caption{DoT - Do53, lossy 4G network}
      \label{fig:seoul_4g_lossy_dot_dns}
  \end{subfigure}
  \hfill
  \begin{subfigure}[t]{0.23\textwidth}
      \includegraphics[width=\linewidth]{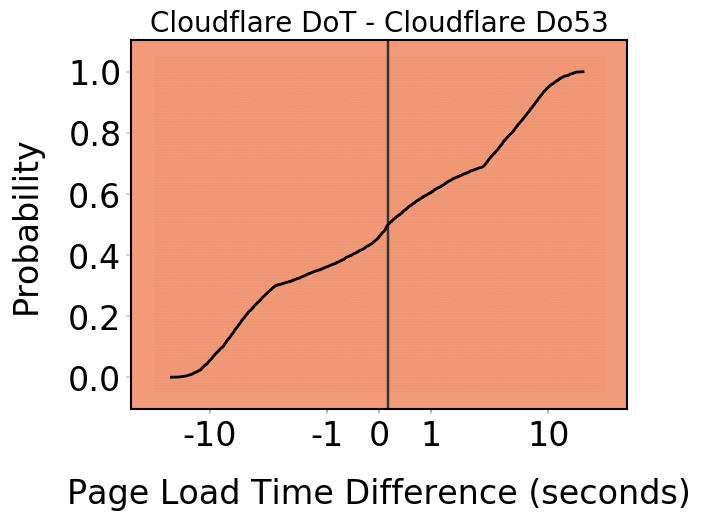}
      \caption{DoT - Do53, 3G network}
      \label{fig:seoul_3g_dot_dns}
  \end{subfigure}
\\
    \vspace{0.5cm}
  \begin{subfigure}[t]{0.23\textwidth}
      \includegraphics[width=\linewidth]{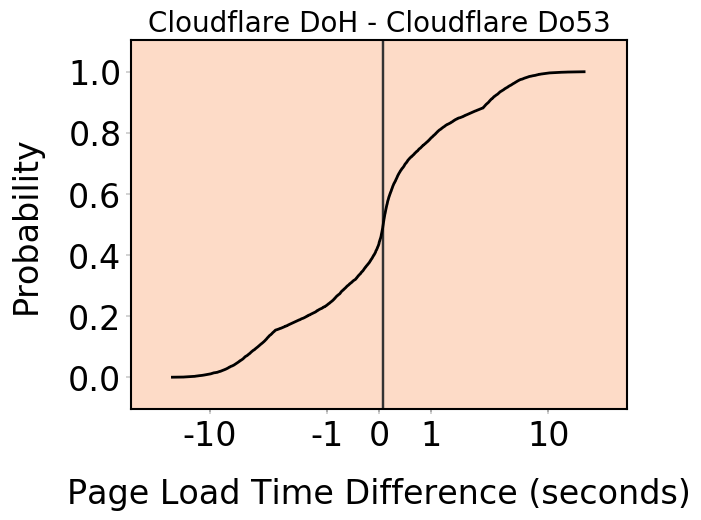}
      \caption{DoH - Do53, Seoul's default network}
      \label{fig:seoul_default_doh_dns}
  \end{subfigure}
  \hfill
  \begin{subfigure}[t]{0.23\textwidth}
      \includegraphics[width=\linewidth]{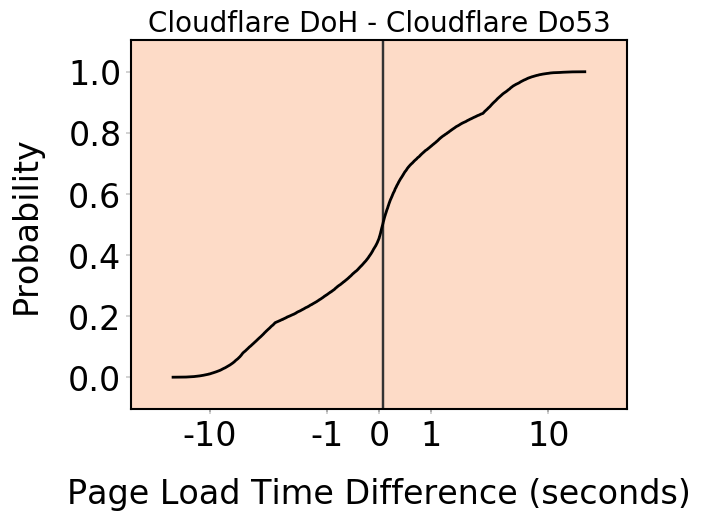}
      \caption{DoH - Do53, 4G network}
      \label{fig:seoul_4g_doh_dns}
  \end{subfigure}
  \hfill
  \begin{subfigure}[t]{0.23\textwidth}
      \includegraphics[width=\linewidth]{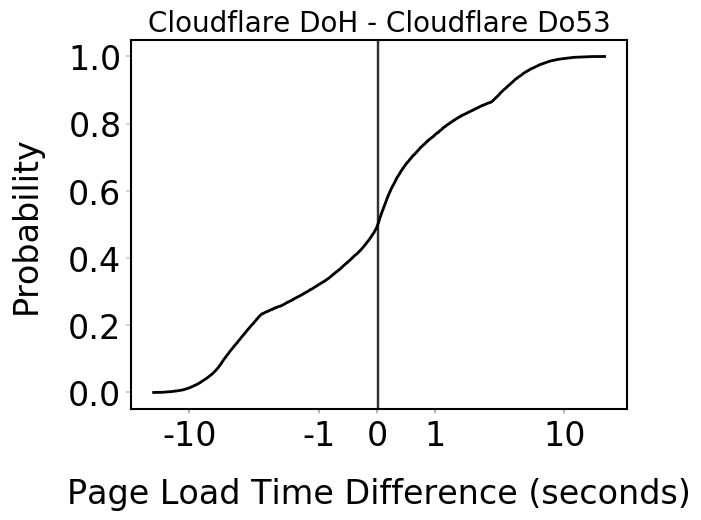}
      \caption{DoH - Do53, lossy 4G network}
      \label{fig:seoul_4g_lossy_doh_dns}
  \end{subfigure}
  \hfill
  \begin{subfigure}[t]{0.23\textwidth}
      \includegraphics[width=\linewidth]{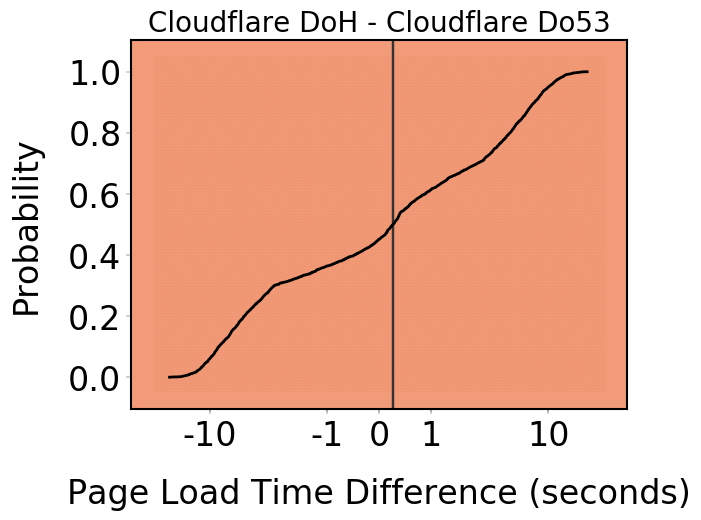}
      \caption{DoH - Do53, 3G network}
      \label{fig:seoul_3g_doh_dns}
  \end{subfigure}
\\
  \begin{subfigure}[t]{\textwidth}
      \includegraphics[width=\linewidth]{figures/pageload_diff_subset_legend}
      \label{fig:cf_pageloads_legend}
  \end{subfigure}
    \caption{Comparison of page load times between protocols and network conditions using Cloudflare's recursors from Seoul}
    \label{fig:seoul_cf_pageloads}
\end{figure*}

Figure~\ref{fig:seoul_cf_pageloads} compared page load times between protocols and network conditions using Cloudflare's recursor from Seoul.
Cloudflare DoT and DoH are slower than Do53 in page load times for the default network condition.
DoT performs 1ms slower than Do53 in the median case, and DoH performs 79ms slower than Do53.
On the 4G network, DoT and DoH performs similarly to how they perform without traffic shaping.
DoT performs 1ms slower than Do53 in the median case, and DoH performs 70ms slower than Do53.

On the lossy 4G network, DoT grows increasingly \emph{faster} than Do53, and DoH begins to close the gap.
DoT performs 45ms faster than Do53 in the median case, and DoH performs 12ms slower than Do53.
As previously discussed, we attribute this improved performance to TCP re-transmitting packets faster than UDP timeouts.
However, page load times with DoT and DoH are both worse than Do53 on an emulated 3G network in Seoul.
DoT performs 175ms slower than Do53 in the median case, and DoH performs 265ms slower than Do53.
Again, we attribute this behavior to DoT and DoH queries contributing to link saturation.

As with Frankfurt, we see that in lossier conditions, DoH experiences higher failure rates compared with Do53.
\footnote{Due to space constraints, we can not include the full failure table for Seoul.}
On the emulated 3G network, Cloudflare Do53 has ${\approx}$21\% less page load timeouts than Cloudflare DoH.
DoT also continues to maintain higher rates of success than DoH, with ${\approx}$21\% less page load timeouts.
Lastly, DNS errors for DoH spike on the emulated 3G network, with ${\approx}$38\% of page loads failing as a result.
We attribute these DNS errors to query timeouts.

The general trend we observe is that page load times with DoT and DoH can improve compared to Do53 in the face of packet loss and high latency.
However, as network conditions degrade, DoT and DoH both perform significantly slower than Do53.
Furthermore, page loads with DoH fail much more often than Do53 and DoT on emulated 3G network conditions.
We note that we are not making a recommendation about which protocol or recursor to use.
We also can not generalize our results to vantage points that we have not measured.
Nonetheless, our results show that your network and choice of DNS transport matter for web performance.

\section{Discussion}\label{sec:discussion}
Based on our results, we offer several insights to improve Do53, DoT, and DoH resolution times, which can reduce page load times and improve user experience.
We first propose opportunistic partial responses, followed by wire-format caching.
We then discuss how dropping support for EDNS Client-Subnet at public recursors may improve page load times.

\subsection{Opportunistic Partial Responses}
\label{sec:discussion:opr}
We discovered that current DNS clients do not utilize part of the DNS Internet Standard that could improve client performance and user experience.
Unfortunately, the three public recursors we measured \emph{violate} the standard~\cite{rfc1035} by \emph{not supporting} queries with more than one question (\(\texttt{QDCOUNT} > 1\)).
Cloudflare and Quad9 do not respond, and Google only responds to the first question.

Without compatible recursors, clients cannot utilize this part of the standard to send fewer larger queries, and, thus, less bytes due to reduced overhead.
We were unable to discover any reason in RFCs and on the IETF dnsop and dnsext mailing lists why servers may misbehave.
We speculate that it could be because the DNS Internet Standard sets the expectation that \texttt{QDCOUNT} is ``usually 1''~\cite{rfc1035}.

Na\"ively, it appears that there is no reason to support more than one question because it would delay the response to a query until all answers have been received, which may take multiple seconds and, in turn, severely degrade user experience.
Furthermore, it would effectively eliminate the benefit of out of order responses that single question queries enable.
Out of order responses are currently implemented in Do53 through UDP, in DoT through response reordering~\cite{rfc7766}, and in DoH through HTTP/2's stream multiplexing~\cite{rfc7540}.

We believe that \emph{opportunistic partial responses} could be a solution:
A client indicates that it wants to use partial responses on the first single question query through a EDNS partial response option, and the server confirms if it supports it.
The client can then send multiple questions in the same query when with the EDNS partial response option, and the server can respond with individual or multiple answers in a DNS response as authoritative answers arrive.
We are currently exploring authoring a corresponding Internet-Draft.

\subsection{Wire Format Caching}\label{sec:wire_format}
Over the course of measurements, we found that Firefox uses a hard-coded DNS transaction ID of 0 for its DoH implementation~\cite{doh-firefox-hardcoded}, which we also use in our query measurement tool.
We posit that this could enable DoH recursors to leverage HTTP response caching of the DNS response's wire format more aggressively and at the edge.
By fixing a transaction ID at the client, recursors could side-step the issue of always having to construct a DNS response, instead reading the wire-format from a local HTTP cache.

The security effect of a fixed transaction ID is limited for DoH because it relies on TLS, which makes it difficult to inject a spoofed response that could be used to poison the client's cache.
For DoT, the same argument can be made and it is similarly amenable to wire format caching.
For Do53, a fixed transaction ID would allow cache poisoning, and, hence, it is not a viable solution.

Generally, to improve tail response times, we suggest to cache the DNS response wire format regardless of transaction ID, and to simply replace the two byte transaction ID before responding (e.g., via \texttt{XOR}), which also has the benefit of being compatible with DoT clients that send random transaction IDs.
It is important to note that the DNS TTL values of a response also need to be updated (decremented) regularly, and this invalidates the HTTP response or wire format cache, but by decreasing the TTL by more than the required amount, the wire format cache can be kept valid longer.

\subsection{EDNS Client-Subnet}\label{sec:recursor}
Cloudflare's recursors result in consistently lower page load times than any other recursor we measured, including the default Do53 recursor provided by Amazon in Frankfurt (\Fref{fig:frankfurt_pageload_diff}, H1 through J10).
We posit that Cloudflare's caching strategy is a core reason for their better performance.
Specifically, their recursors can cache responses more easily because they do not support EDNS Client-Subnet(ECS)~\cite{edns-cloudflare,rfc7871}, which Google generally supports~\cite{edns-google}. 

The purpose of ECS is to forward the client's address or network to the authoritative server via the recursor, which allows the authoritative server to provide a response to the recursor that takes the client's address into account, for example to direct it to a server that is located nearby.
By not supporting ECS, Cloudflare's recursors can have higher cache hit rates, in particular for a client's first queries.
Specifically, Cloudflare does not need to limit cached responses to the client's IP address or network indicated through ECS in the original query, that is, their cache is client agnostic.
On the contrary, the caches for Google and partially Quad9 must be client specific because of ECS.

Website and CDN operators should therefore consider abandoning DNS-based localization and stop relying on ECS, and instead adopt anycast.
Interestingly, the cost that recursor cache misses incur because of ECS could actually negate the benefits of directing a user to a local server via ECS in a variety of cases, and even directing him to a single central data center (without anycast) could lead to a better user experience than ECS.
Overall, \emph{disabling ECS not only improves client privacy, but our results show that it may also decrease client page load times, leading to an immediate improvement in a user's browsing experience}.

\section{Related Work}\label{sec:related}
In this section, we first compare to related work on DNS privacy and security.
We then compare to measurements on how DNS impacts web performance.

\subsection{Encrypted DNS Transports}
Zhu et al.~\cite{zhu2015connection} introduced DNS over TLS, that is DNS over TLS over TCP, to provide confidentiality guarantees that DNS lacked.
They measured the performance costs and benefits of sending DNS queries over a TLS connection, and find that DoT response times are only up 22\% slower than Do53.
We measure higher DoT response times when measuring response times na\"ively due to fewer queries being sent and less connection reuse.
Different from Zhu et al., our study focuses on how different DNS transports affect user experience, through page load times, and how it differs in the face of different network conditions.

Böttger et al. measured query response times and page load times for Do53, DoT, and DoH from a university network~\cite{boettger2019empirical}.
Unfortunately, their methodology relies on collecting HARs for query response time measurements.
As we discuss in~\ref{sec:dns_method}, HARs can contain invalid response times depending on how re-directs are triggered.
This is also evident from Figure 6 in their paper showing a y-intercept of approximately 10\%, which means that for roughly 10\% of websites the DNS resolution for all included resources can be performed sequentially in 0ms.

In addition to DoT and DoH, other protocols have been proposed to help ensure privacy and security between a client and a recursor.
DNSCrypt utilizes cryptographic signatures to authenticate a recursor to a client, which prevents DNS responses from being spoofed or tampered with~\cite{dnscrypt}.
DNSCurve utilizes elliptic-curve cryptography to provide confidentiality, authenticity, and integrity of DNS responses~\cite{dnscurve}.
However, for DNSCrypt, DNSCurve, DoT, and DoH, the recursor remains aware of what names a client queries for, which has privacy implications as it allows the recursor to learn about the websites that the client visits and when it visits them.
Schmitt et al.~\cite{schmitt2019oblivious} proposed Oblivious DNS, which prevents a recursor from associating queries to the clients that sent them.
This in turn prevents a recursor from learning the client's browsing history.

\subsection{DNS and Web Performance}
Sundaresan et al.~\cite{sundaresan2013measuring} measured and identified performance bottlenecks for web page load time in broadband access networks and found that page load times are influenced by slow DNS response times and can be improved by prefetching.
An important distinction is that they define the DNS response time only as the response time for the first domain, while we consider the set of unique fully qualified domain names of all resources contained in a page.
They investigate only nine high-profile websites, which stands in contrast to the 2,000 popular and normal websites that we analyze, and they estimate page load times through Mirage and validate their findings through a headless browser PhantomJS, while we utilize Mozilla Firefox, which is a full browser.
Wang et al.~\cite{wang2013demystifying} introduced WProf, which is a profiling system to analyze page load performance.
They identified that DNS queries--in particular uncached, cold queries--can significantly affect web performance, accounting for up to 13\% of the critical path delay for page load times.

In 2012, Otto et al.~\cite{otto2012content} found that CDN performance was negatively affected when clients choose recursors that were geographically separated from CDN caches.
They conjectured that this poor performance was a result of recursors not supporting ECS.
Indeed, ECS was only introduced in January 2011, and it was not standardized until May 2016~\cite{rfc7871}.
Therefore, clients were likely redirected to sub-optimal data center based on the recursor's address or network, instead of the client's address.
Otto et al. proposed \texttt{namehelp}, a DNS proxy that improves CDN performance for these far away recursors.
It sends DNS queries for CDN-hosted content directly to authoritative servers, enabling CDNs to use the client's IP address.
We suspect that with the wide-spread adoption of ECS and anycast since 2012, CDN performance may not be as negatively affected by choosing a recursor that is geographically far away from a CDN.

\label{lastpage}\section{Conclusion}\label{sec:conclusion}
In this paper, we investigated DNS timings and page load times using different DNS transport protocols, recursors, network conditions, and global vantage points.
We find that although DoT and DoH result in higher response times for individual queries, they can perform similarly to Do53 in page load times.
We also find that DoT and DoH can outperform Do53 in page load times in emulated cellular network conditions.
However, as network conditions degrade, Do53 significantly outperforms DoT and DoH.
Web pages also load successfully more often with Do53 in poor network conditions.

Based on our findings, DNS stakeholders can take several concrete steps to improve query response times, and in turn page load times.
For example, Firefox currently uses synchronous calls for Do53 and DoT resolution, and asynchronous calls could benefit performance.
Another opportunity to improve Do53 and DoT response times that we discovered is wire format caching.
Lastly, clients and recursors could be extended to support multiple questions in a single query and opportunistic partial responses.
This could be accomplished in a backward compatible way through a new EDNS option.

\microtypesetup{protrusion=false}

\balance
\begin{sloppypar}
\printbibliography
\end{sloppypar}

\microtypesetup{protrusion=true}

\end{document}